\documentclass[conference]{IEEEtran}
\IEEEoverridecommandlockouts
\usepackage{cite}
\usepackage{amsmath,amssymb,amsfonts}
\usepackage{algorithmic}
\usepackage{graphicx}
\usepackage{textcomp}
\usepackage{xcolor}

\usepackage{caption}
\usepackage{booktabs}
\usepackage{amsmath}
\usepackage{multirow}
\usepackage{tabularx} 
\usepackage[hidelinks]{hyperref} 

\usepackage{float}

\def\BibTeX{{\rm B\kern-.05em{\sc i\kern-.025em b}\kern-.08em
    T\kern-.1667em\lower.7ex\hbox{E}\kern-.125emX}}
\begin{document}


\title{Art2Music: Generating Music for Art Images with Multi-modal Feeling Alignment\\}

\author{\IEEEauthorblockN{Jiaying Hong}
\IEEEauthorblockA{\textit{School of Computing} \\
\textit{Newcastle University}\\
Newcastle upon Tyne, UK \\
hongjalynn@gmail.com}
\and
\IEEEauthorblockN{Ting Zhu}
\IEEEauthorblockA{\textit{School of Computing} \\
\textit{Newcastle University}\\
Newcastle upon Tyne, UK \\
t.zhu11@newcastle.ac.uk}
\and
\IEEEauthorblockN{Thanet Markchom}
\IEEEauthorblockA{\textit{Department of Computer Science} \\
\textit{University of Reading}\\
Reading, UK \\
thanet.markchom@reading.ac.uk}
\and
\IEEEauthorblockN{Huizhi Liang}
\IEEEauthorblockA{\textit{School of Computing} \\
\textit{Newcastle University}\\
Newcastle upon Tyne, UK \\
huizhi.liang@newcastle.ac.uk}
}

\maketitle

\begin{abstract}

With the rise of AI-generated content (AIGC), generating perceptually natural and feeling-aligned music from multimodal inputs has become a central challenge. Existing approaches often rely on explicit emotion labels that require costly annotation, underscoring the need for more flexible feeling-aligned methods. To support multimodal music generation, we construct \textbf{ArtiCaps}, a pseudo feeling-aligned image–music–text dataset created by semantically matching descriptions from ArtEmis and MusicCaps. We further propose $Art2Music$, a lightweight cross-modal framework that synthesizes music from artistic images and user comments. In the first stage, images and text are encoded with OpenCLIP and fused using a gated residual module; the fused representation is decoded by a bidirectional LSTM into Mel-spectrograms with a frequency-weighted L1 loss to enhance high-frequency fidelity. In the second stage, a fine-tuned HiFi-GAN vocoder reconstructs high-quality audio waveforms. Experiments on \textbf{ArtiCaps} show clear improvements in Mel-Cepstral Distortion, Fréchet Audio Distance, Log-Spectral Distance, and cosine similarity. A small LLM-based rating study further verifies consistent cross-modal feeling alignment and offers interpretable explanations of matches and mismatches across modalities. These results demonstrate improved perceptual naturalness, spectral fidelity, and semantic consistency. Art2Music also maintains robust performance with only 50k training samples, providing a scalable solution for feeling-aligned creative audio generation in interactive art, personalized soundscapes, and digital art exhibitions.

\end{abstract}


\section{Introduction}
In recent years, the rapid advancement of AI-Generated Content (AIGC) technologies has positioned multimodal generative models at the core of intelligent content creation. The integration of heterogeneous modalities such as images, text, and audio has become a key direction for promoting Web intelligence, context-aware interaction, and immersive content generation. Images (i.e., paintings) and music are two of the most popular forms of artwork, both capable of evoking deep and similar feelings in people. Different artistic modalities can resonate with one another by conveying comparable emotional undertones. For example, Arnold Böcklin’s painting Die Toteninsel inspired Rachmaninoff’s symphonic poem Isle Of The Dead, both evoking a profound sense of mystery and mortality~\cite{b21}. However, generating feeling-aligned and perceptually natural audio from artistic images and their accompanying textual commentary remains an emerging research challenge.

MusFlow~\cite{b1} and Mozart’s Touch~\cite{b2} proposed frameworks that incorporate visual and linguistic modalities to guide music generation, demonstrating the potential of multimodal modeling in the field of artistic creation. The work~\cite{b3} goes a step further by attempting to map emotional dimensions in paintings to musical styles, emphasizing the importance of emotional consistency. However, most existing research faces three key limitations.


First, existing studies primarily focus on emotion-driven music generation, whereas our work extends this perspective to the broader concept of feeling, which encompasses—but is not limited to—both emotional and feeling dimensions. Second, existing approaches generally rely on large-scale Transformer architectures \cite{b4}, which are computationally expensive and lack adaptability for lightweight deployment in low-resource settings. Lastly, constructing high-quality image-text-audio triplet datasets typically requires extensive manual annotation, resulting in significant resource consumption and limiting the scalability and generalizability of related methods.


To address the first limitation, we define \textit{feeling} as a holistic perception that goes beyond narrow notions of emotion. It not only includes the basic emotional responses elicited by artworks but also extends to the atmosphere, subjective experience, and cross-sensory qualities shaped by visual and auditory modalities, such as a sense of nostalgia, solemnity, or solitude, as well as perceptual impressions like the warmth or coolness of colors and the dynamism of visual lines. To make this concept computationally tractable in our framework, \textit{feeling} is operationalized as a multimodal representation learned through OpenCLIP~\cite{b5,b6}. Visual and textual modalities are projected into a shared latent space that jointly encodes affective tone as well as macro- and micro-level perceptual attributes such as atmosphere, warmth, or texture. The fused embedding obtained through the gated residual fusion module thereby forms a feeling space, establishing a continuous and reproducible mapping between visual–text semantics and cross-modal perceptual experience.

To tackle the second limitation, we propose a lightweight multimodal audio generation framework Art2Music~\footnote{\url{https://github.com/hh-jy/Art2Music/tree/main}}, aimed at extracting deep semantic representations from artistic images and textual commentary, and generating audio that is feeling-aligned with the visual-linguistic content. Art2Music employs a two-stage architecture comprising mel-spectrogram generation and audio reconstruction. It performs modality alignment using OpenCLIP encoders and a residual fusion module, and generates high-quality audio outputs through a Long short-term memory (LSTM)~\cite{b7} decoder and a HiFi-GAN vocoder~\cite{b8}.

To address the scarcity of tri-modal paired data, we propose a weak alignment strategy based on semantic similarity to match two heterogeneous data sources, ArtEmis~\cite{b9} (which includes artworks and commentary text) and MusicCaps~\cite{b10} (which contains music and descriptive texts). This dataset is publicly available to facilitate multi-modal music generation tasks. Furthermore, we introduce a frequency-aware loss function to emphasize high-frequency components during spectrogram reconstruction, enhancing audio fidelity and expressiveness. The key contributions of this work can be summarized as follows:

\begin{itemize}
    \item A lightweight, feeling-aligned cross-modal music generation framework that synthesizes high-quality music from artistic images and user text comments or descriptions.

    \item A new feeling-aligned multi-modal dataset ArtiCaps that aligns artistic images and music with common feelings.


    \item A frequency-weighted L1 loss is designed to prioritize high-frequency regions in Mel-spectrogram reconstruction, improving perceptual fidelity in generated audio.

    \item Extensive experiments and rigorous evaluations conducted on ArtiCaps dataset demonstrate high feeling consistency and diverse audio outputs under limited data and computational resources.    

\end{itemize}

\section{Related Work}
\subsection{Multimodal Music Generation}
In recent years, the rapid progress in multimodal generation has enabled audio synthesis from multiple modalities such as text, images, and emotional labels. MusFlow~\cite{b1} proposed a music generation framework based on conditional flow matching, incorporating image and text inputs, and introduced the multimodal dataset MMusSet to support modality alignment and conditional generation. Mozart's Touch~\cite{b2} designed a lightweight framework utilizing large language models (LLMs), but still relies on pre-trained components such as Bootstrapping Language-Image Pre-training for Unified Vision-Language Understanding and Generation (BLIP) and MusicGen. M\textsuperscript{2}UGen~\cite{b12} further presented a unified framework to generate from image, text, and audio modalities through LLM-based cross-modal understanding and generation, demonstrating strong generalization. However, these methods are mostly built upon LLMs, lacking lightweight and semantically driven modeling capabilities, and often require high inference complexity and resource cost.

In addition, \cite{b3} explored the mapping of emotional features extracted from paintings into musical styles, highlighting the importance of emotional consistency in generative music. Overall, existing approaches primarily focus on emotion-guided or LLM-based strategies, while systematic exploration of lightweight, feeling-aligned modeling beyond explicit emotion labels remains limited.

Existing open-source models (e.g., MusicLM~\cite{b10}, MusicGen~\cite{b20}) remain limited to single-modality inputs, while M\textsuperscript{2}UGen leverages closed-source pipelines, hindering reproducibility. Accordingly, we conduct evaluations across different input conditions using established objective metrics and complementary LLM-based subjective assessments.

\subsection{Modality Alignment and Fusion}
Maintaining perceptual or stylistic consistency has been extensively studied in visual generative models. For instance, Ko \textit{et al.}~\cite{b24} proposed a GAN-based framework for Korean font synthesis that preserves perceptual style coherence across generated glyphs. Such visual consistency methods highlight the broader importance of alignment in generative modeling, which extends naturally to cross-modal scenarios.

Effective alignment between vision and language modalities is a key prerequisite for high-quality multimodal generation. Recent advances such as Contrastive Language-Image Pre-Training (CLIP)~\cite{b5} and OpenCLIP~\cite{b6} have been widely applied in joint vision-language representation learning. Compared to CLIP, OpenCLIP offers greater flexibility in text encoding and stability, making it more suitable for complex language input scenarios and thus broadly adopted in multimodal retrieval and generation tasks.

For modality fusion, various strategies have been explored to integrate cross-modal features, including vector concatenation~\cite{b2}, multimodal compact bilinear pooling (MCB)~\cite{b13}, and attention-based cross-modal interaction mechanisms~\cite{b14}. While these methods enhance modality interaction, they introduce substantial model complexity and computational cost, limiting their applicability to lightweight or edge-device deployments.

At the data level, the scarcity of paired image-text-audio datasets remains a major bottleneck in building cross-modal generative systems. MMusSet~\cite{b1} constructed in MusFlow is one of the few publicly available tri-modal resources. In addition, weakly supervised semantic matching strategies have been employed in image-text retrieval~\cite{b15}, typically using pre-trained language models such as TinyBERT~\cite{b11} to encode semantic embeddings and construct pseudo-aligned pairs based on similarity. Such approaches provide useful insights for constructing training samples in cross-modal generation, and can be extended to tri-modal settings by inspiring pseudo-aligned triplet construction that connects images, text, and audio.

\subsection{Audio Generation and Mel-to-Waveform Models}
In music and speech synthesis tasks, generating audio from intermediate representations such as Mel-spectrograms is one of the mainstream approaches. HiFi-GAN~\cite{b8} is an efficient neural vocoder capable of producing high-fidelity waveforms from spectrograms with strong real-time synthesis performance, and has been widely adopted in text-to-speech (TTS)~\cite{b16} and music generation scenarios.

By contrast, models such as AudioLM~\cite{b17} and MusicLM\cite{b10} employ end-to-end audio modeling strategies, generating audio from text input via discretized audio representations and language models. While these approaches exhibit excellent modeling capacity, they typically rely on complex multi-stage pretraining and require substantial hardware resources. In comparison, two-stage approaches based on spectrogram modeling strike a balance between generation quality and efficiency, making them more suitable for deployment in resource-constrained generation systems. In terms of sequence modeling architectures, LSTM networks~\cite{b7} have been widely adopted as classical structures for speech and audio sequence prediction in earlier studies. Models such as SampleRNN~\cite{b18} demonstrated strong generative capabilities based on LSTM architectures. Although Transformer-based models have become dominant in recent years, LSTMs remain suitable for lightweight modeling scenarios due to their smaller parameter size and stable convergence properties.

Despite progress, three gaps remain: (1) lack of lightweight frameworks that do not rely on large language models; (2) overreliance on explicit emotion labels, with limited study of feeling-aligned modeling; and (3) scarcity of tri-modal datasets to support cross-modal training. These gaps call for a lightweight, feeling-aligned framework for efficient and reproducible multimodal generation. Our proposed Art2Music directly bridge the gap between efficiency, perceptual alignment, and multimodal scalability.

\section{The Proposed Approach}
In this section, we first introduce the construction of the dataset, followed by our proposed feeling-aligned cross-modal music generation framework.

\subsection{Dataset Construction and Representation}

\begin{table}[t]
\scriptsize
\centering
\setlength{\tabcolsep}{3pt}
\caption{Example of a tri-modal sample with left-column image, middle-column textual fields, and right-column audio (icon with link).}
\label{tab:articaps_sample}
\begin{tabular}{@{}p{2.5cm} p{3.8cm} p{2.0cm}@{}}
\toprule
\multirow{5}{*}{\raisebox{-1.25\height}{\includegraphics[width=2.5cm]{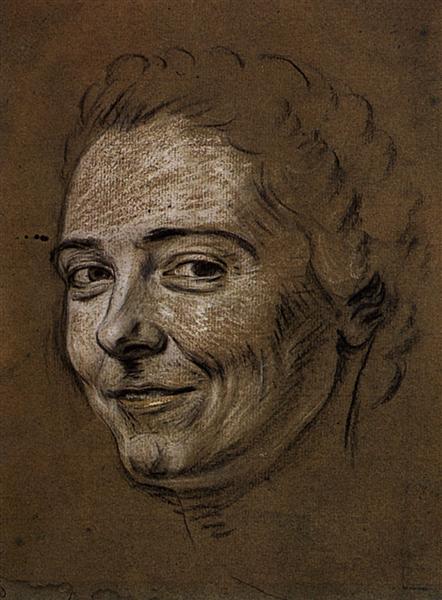}}}
& \textbf{Art Commentary:} Looks like a painting of some smug weirdo
& \multirow{5}{*}{\raisebox{-2.2\height}{%
    \href{https://www.youtube.com/watch?v=MX0wS7MX3Zo}
    {\begin{minipage}{2.0cm}
      \centering
      \textbf{Matched Audio (1:30-2:30 minutes)}\\[2pt]
      \includegraphics[width=0.7\linewidth]{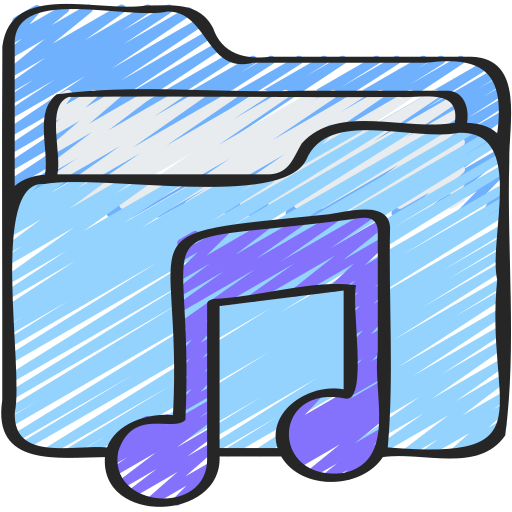}
    \end{minipage}}
}} \\ \cmidrule(l){2-2}
& \textbf{Painting Emotion:} smug \\ \cmidrule(l){2-2}
& \textbf{Audio Caption Keywords:} classical music, contemporary, theremin, electronic sounding, instrumental, weird, eerie, eccentric \\ \cmidrule(l){2-2}
& \textbf{Audio Set Label:} Music,Electronic music,House music,Trance music \\ \cmidrule(l){2-2}
& \textbf{Audio Caption:} This is a contemporary classic music piece being performed on a theremin. The player gives the theremin and electronic sounding character. The atmosphere is weird and eccentric. This piece could go well in the soundtrack of an absurdist/surrealist art movie. \\
\bottomrule
\end{tabular}
\end{table}

\subsubsection{Data Source}
The ArtiCaps dataset is built by combining ArtEmis~\cite{b9} and MusicCaps~\cite{b10}. ArtEmis, based on WikiArt, includes about 80,000 artworks and 455,000 user commentaries with emotional attributions. MusicCaps provides 5,521 musician-written audio captions; after cleaning, 4,717 ten-second clips with descriptions are retained.

\subsubsection{Weakly Aligned Triplet Construction} Given the absence of native triplet datasets containing aligned image, text, and audio modalities, we adopt a weak alignment strategy to construct a pseudo-paired dataset named ArtiCaps. A sample of ArtiCaps is shown in Table~\ref{tab:articaps_sample}. We extract images and their commentary texts (Art Commentary) from the ArtEmis dataset and retrieve audio samples and descriptive captions (Audio Caption) from MusicCaps. Both types of text are encoded using TinyBERT~\cite{b11}, and semantic cosine similarity (a metric widely used to measure similarity in semantic embedding spaces) is computed to match each ArtEmis sample with the most semantically similar MusicCaps entry. These matched triplets serve as multimodal training inputs. This approach makes use of the existing user commentary text that contains rich feeling information and does not require extra human annotation.
\subsubsection{Text Preprocessing and Construction} To enrich the semantic input, in addition to the original commentary, we extract emotional keywords from ArtEmis descriptions. We employ the Opinion Lexicon sentiment lexicon~\cite{b19} combined with part-of-speech filtering (adjectives, adverbs, nouns, verbs), retaining only high-emotion-density keywords. These keywords are then augmented with the original artwork emotion labels from ArtEmis to form a richer set of emotional keywords. A composite textual input is then constructed in the following format:
\begin{quote}
"\texttt{Art Commentary: <Art commentary>. Painting Emotion: <Painting Emotion>. Audio: <Audio Set Label>. Audio keywords: <Audio Caption Keywords>.}"
\end{quote}
\textit{\texttt{<Art Commentary>}} contains both the content description of the artwork and the viewer’s emotional response. \texttt{<Painting Emotion>} refers to emotional keywords extracted from the Art Commentary. \texttt{<Audio Set Label>} denotes the music category, and \texttt{<Audio Caption Keywords>} are keywords describing some aspects of music.
\subsubsection{Feeling Consistency of Semantic Matching}

\begin{figure}[t]
\centerline{\includegraphics[width=0.85\linewidth]{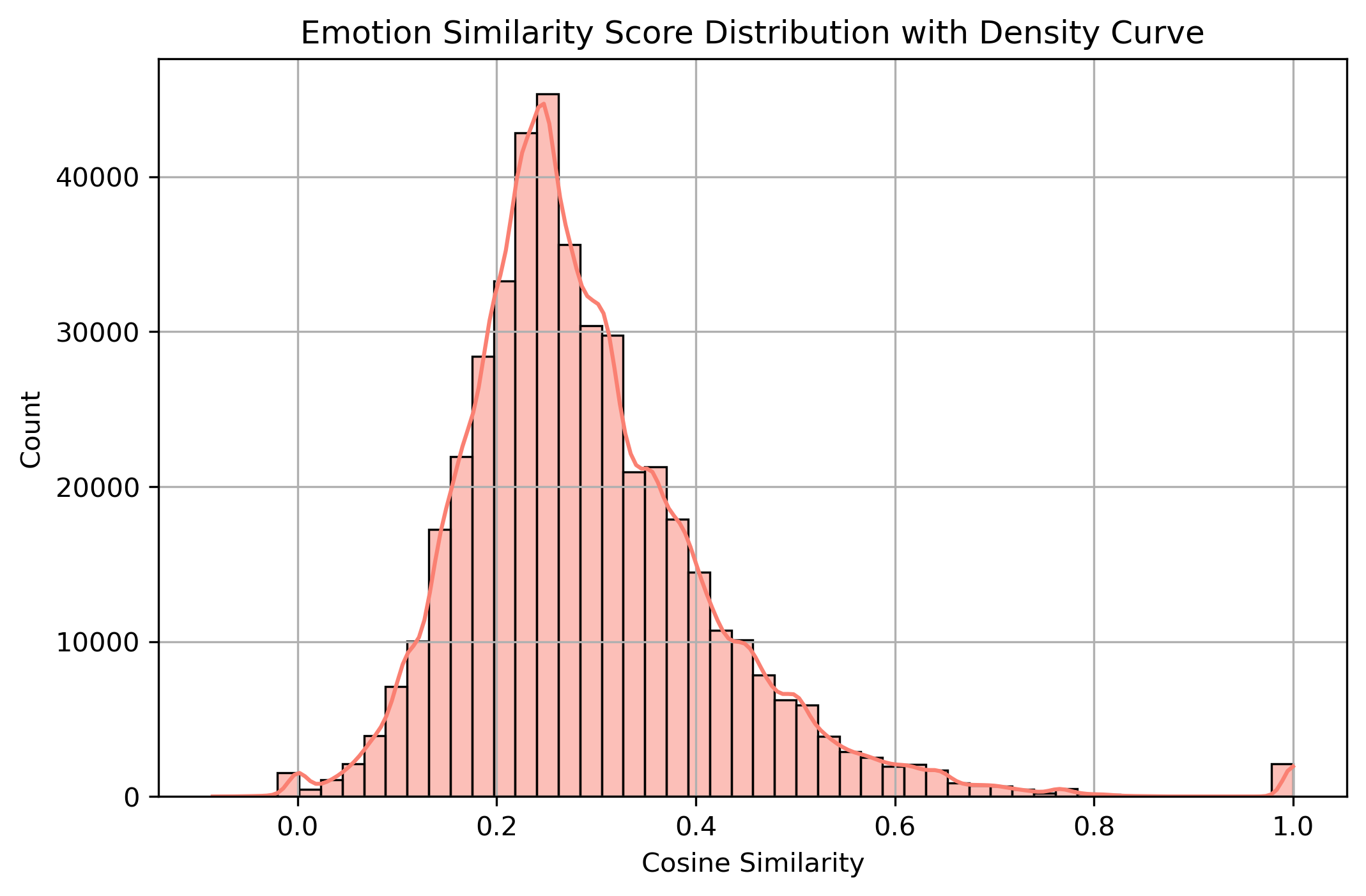}}
\caption{Distribution of cosine similarity scores between painting-side and audio-side emotional keywords.}
\label{fig:similarity_distribution}
\end{figure}

\begin{figure}[htbp]
\centerline{\includegraphics[width=0.85\linewidth]{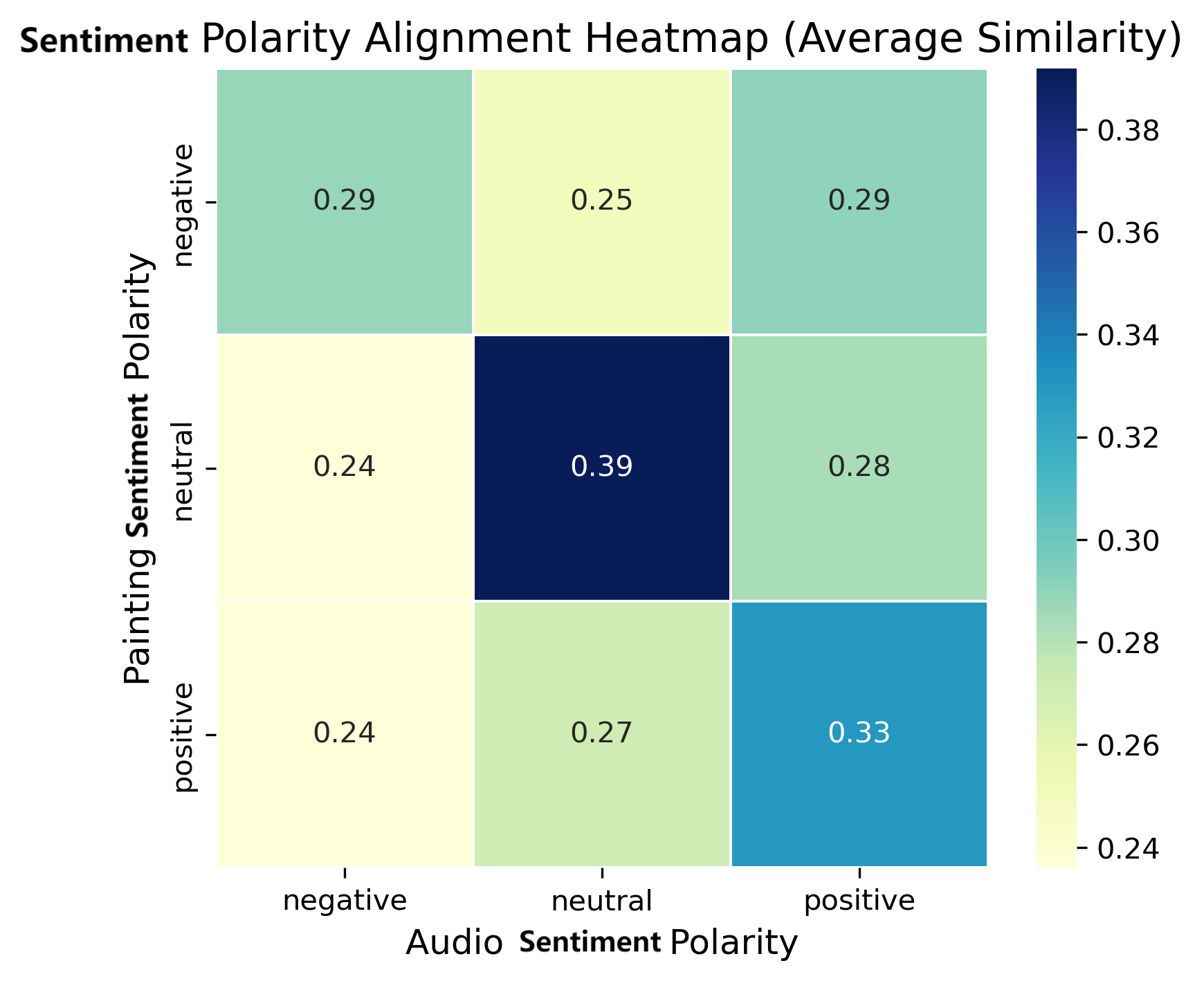}}
\caption{Emotion polarity alignment heatmap showing the average similarity between sentiment polarities of painting and audio modalities. Diagonal entries (e.g., positive–positive, neutral–neutral) exhibit higher values, suggesting retained polarity consistency during weakly supervised semantic matching.}
\label{fig:polarity_heatmap}
\end{figure}

\begin{table*}[!t]
\renewcommand{\arraystretch}{1.8}
\setlength{\tabcolsep}{5pt}
\caption{Examples of weakly aligned text pairs from ArtEmis (visual domain) and MusicCaps (audio domain) with emotional similarity analysis.}
\centering
\small
\resizebox{\textwidth}{!}{%
\begin{tabular}{p{3.3cm} p{3.3cm} p{2.5cm} p{2.5cm} c c c p{3.3cm}}
\toprule
\textbf{Art Comment} & \textbf{Music Description} & \textbf{Painting Emotion} & \textbf{Audio Emotion} & \textbf{Similarity} & \textbf{Paint Polarity} & \textbf{Audio Polarity} & \textbf{Observation} \\
\midrule
The girls look happy and the ducks do as well. The scene is calming and peaceful. & Here we have a slow piano piece played in a major key. The peace feels calm and happy. & contentment, happy, calm, peaceful, well & slow, calm, happy, peace & 0.84 & positive & positive & Lexical differences exist, but the emotional tone is consistent. \\[4pt]
This photo makes me sad because there is a baby in it that appears to be hanging by a tree, and that's sad. The people also look scared. & In this clip, a large bell is rung and left to ring. We can hear the resonance in the room as the bell rings. There is then the faint sound of a male speaking in the background. It's a live recording. & sadness, hang, sad, scared & faint & 0.32 & negative & negative & Low lexical match, but emotional direction remains close. \\[4pt]
Pleased to meet you, which way to the kitchen. & Church bells ringing together slowly. & excitement, pleased & slowly & 0.27 & positive & negative & Audio side lacks emotional cues; similarity score is low. \\
\bottomrule
\end{tabular}
}
\label{tab:emotion_case}
\end{table*}

\begin{table}[t]
\centering
\renewcommand{\arraystretch}{1.25}
\caption{Sentiment Polarity Distribution}
\begin{tabular}{lccc}
\hline
\textbf{Modality} & \textbf{Positive (\%)} & \textbf{Neutral (\%)} & \textbf{Negative (\%)} \\ 
\hline
Painting-side & 55 & 13 & 32 \\
Audio-side & 49 & 17 & 33 \\ 
\hline
\end{tabular}
\label{tab:polarity}
\end{table}

To analyze the feeling alignment characteristics of semantically matched samples, we extract emotional keywords from the textual descriptions on both the visual (painting) and auditory (music) sides. These keywords are encoded using the all-MiniLM-L6-v2 model, and their semantic cosine similarity is computed to assess cross-modal emotional consistency. As shown in Fig.~\ref{fig:similarity_distribution}, most samples fall within the similarity range of 0.2–0.4. Although overall feeling consistency is limited, a subset of samples exhibits strong emotional correspondence. Further, we categorize the extracted keywords into three sentiment polarities: positive, neutral, and negative. We then compute the average similarity scores for all combinations of polarity pairs. As illustrated in Fig.~\ref{fig:polarity_heatmap}, the heatmap reveals that diagonal combinations tend to yield higher similarity scores. This indicates that semantic matching preserves a certain degree of sentiment polarity consistency. Table~\ref{tab:emotion_case} presents several representative aligned sample pairs, including painting-side comments and corresponding audio descriptions, their associated emotional keywords and cosine similarity scores, as well as brief annotations regarding their match quality. We observe that high-scoring pairs often share highly consistent emotional keywords. In contrast, some low-scoring samples exhibit weaker alignment due to audio descriptions focusing more on structural or acoustic properties rather than explicit emotional content. To quantitatively analyze the emotional polarity consistency between modalities, we computed the sentiment distribution of both painting- and audio-side texts based on their emotion labels. As shown in the Table~\ref{tab:polarity}, for painting-side texts, the sentiment polarity distribution was 55\% positive, 13\% neutral, and 32\% negative, while for audio-side texts it was 49\% positive, 17\% neutral, and 33\% negative, showing overall balance between modalities. Across all matched pairs, 47\% shared the same sentiment polarity, whereas 42\% had cosine similarity values below 0.25. This moderate level of alignment reflects the weakly aligned matching strategy of ArtiCaps with respect to emotion-focused sentiment, which is limited to three generic emotional categories. However, the proposed method is designed to capture broader feeling correspondence rather than strict emotional equivalence. Despite these variations, the matched samples collectively reveal stable perceptual tendencies that are consistent at the overall tone or feeling level. Nevertheless, these samples still display perceptible consistency in overall tone or feeling atmosphere. This analysis of emotional consistency reinforces the interpretability of our multimodal alignment strategy and provides a conceptual basis for future generative models that integrate feeling components.
\subsubsection{Audio to Mel-Spectrogram Conversion} All audio samples are resampled to 22.05\,kHz (a common choice in music generation, balancing fidelity and efficiency while matching HiFi-GAN vocoder settings). Mel-spectrograms are extracted using Librosa~\cite{b22} with standard parameters. The FFT window size is set to 1024 (\texttt{FFT=1024}), the hop length to 256 (\texttt{hop=256}), and the number of Mel bands to 80 (\texttt{$n_{\text{mels}}$=80}). These parameters respectively determine the frequency resolution, the time resolution, and the number of frequency channels in the spectrogram. The extracted spectrograms are then normalized to the range [-1, 1]. To ensure consistent input dimensions during both training and inference, all spectrograms are adjusted to a fixed time length (\texttt{T=896}); shorter sequences are zero-padded, and longer ones are truncated. The processed spectrograms serve as learning targets for the Mel decoder~\cite{b8}.

\subsection{The Proposed Art2Music framework}

\begin{figure*}[t]
\centering
\includegraphics[width=\textwidth, height=8cm]{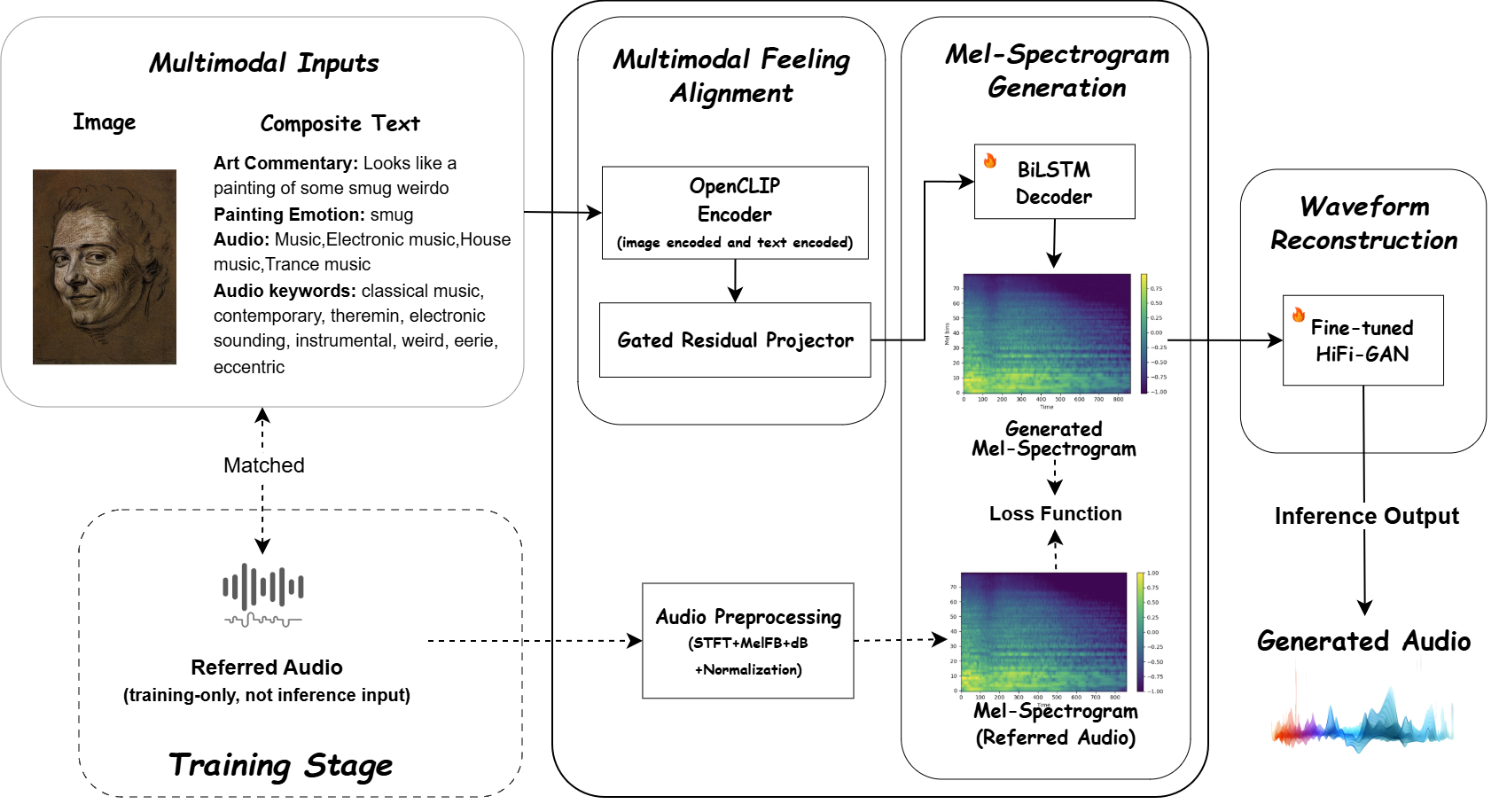}
\captionsetup{width=\textwidth}
\caption{Overview of the proposed Art2Music framework, consisting of two stages: (1) multimodal feeling alignment and Mel-spectrogram generation from image and textual input; (2) waveform reconstruction using a HiFi-GAN vocoder.}
\label{fig:framework}
\end{figure*}

The framework is shown in Fig.~\ref{fig:framework}. It is designed to synthesize audio waveforms that align with input images (artworks) and textual descriptions (including artwork commentary and audio cues). The overall pipeline consists of two sequential stages: (1) \textbf{Multimodal semantic encoding and spectrogram generation}, which transforms aligned image-text into Mel-spectrograms; (2) \textbf{Mel-based waveform reconstruction}, where spectrograms are decoded into high-quality playable audio via HiFi-GAN vocoder. This modular design supports efficient inference, flexible training, and better adaptability to low-resource multimodal generation scenarios.
\subsubsection{Multimodal Feeling Alignment}
The input image is encoded into a visual feature vector using OpenCLIP (ViT-H/14)~\cite{b6}, while the composite textual prompt is embedded using OpenCLIP’s text encoder~\cite{b6}. To bridge the semantic gap between modalities and enable cross-modal interaction, we introduce a Gated Residual Projector.  This module concatenates the image and text embeddings, projects them into a shared latent space, and retains a residual path from the text feature. A sigmoid-based gating mechanism adaptively balances the contribution of each component. The computation is as follows:
\begin{equation}
\mathbf{h} = \sigma(\mathbf{W}_g [\mathbf{x}; \mathbf{r}]) \odot \mathbf{W}_x \mathbf{x} + (1 - \sigma(\mathbf{W}_g [\mathbf{x}; \mathbf{r}])) \odot \mathbf{r}
\end{equation}
where $\mathbf{x}$ is the fused image-text representation, $\mathbf{W}_g$ and $\mathbf{W}_x$ are learnable projection matrices, $\mathbf{r}$ is the residual text embedding, $\sigma$ denotes the sigmoid function, and $\odot$ represents element-wise multiplication. This lightweight module enables fine-grained alignment between image and text representations in the shared latent space, while preserving computational efficiency.

\subsubsection{Mel-Spectrogram Generation Module}
The aligned semantic representation $\mathbf{h}$ is projected into a sequence format and encoded with positional embeddings. We employ a 4-layer bidirectional LSTM decoder to generate Mel-spectrograms of fixed resolution (\texttt{T=896}, \texttt{$n_{\text{mels}}$=80}). LSTM is chosen for its lightweight architecture, temporal modeling capacity, and stable convergence in low-resource scenarios.
To emphasize high-frequency detail and audio fidelity, we design a frequency-weighted L1 loss function, defined as:

\begin{equation}
\mathcal{L}_{\text{mel}} = \frac{1}{TF} \sum_{t=1}^{T} \sum_{f=1}^{F} w_f \cdot \left| \hat{M}_{t,f} - M_{t,f} \right|
\end{equation}
where $\hat{M}$ and $M$ denote the predicted and reference Mel-spectrograms respectively. $T$ is the number of time frames, and $F$ is the number of Mel frequency bins. The weight vector $w = [w_1, \dots, w_F] \in \mathbb{R}^F$ is predefined such that $w_f$ increases linearly from $1.0$ at the lowest frequency bin to $1.5$ at the highest one, i.e., $w_f = 1.0 + 0.5 \cdot \frac{f-1}{F-1}$. We set the weighting range to 1.0–1.5 to retain low-frequency fidelity (weight = 1.0) while providing a mild high-frequency emphasis (up to 1.5), as larger values risk over-biasing the optimization toward high frequencies and destabilizing training. This explicit design ensures that errors in higher-frequency bins are assigned larger weights, 
thereby encouraging the model to focus more on high-frequency details that are harder to synthesize but critical for perceptual quality. 

\subsubsection{Waveform Reconstruction Module}
The predicted Mel-spectrograms are converted into time-domain waveforms using a fine-tuned HiFi-GAN vocoder~\cite{b8}. HiFi-GAN offers high-fidelity, low-latency, and real-time synthesis capabilities. It supports effective reconstruction under variable acoustic conditions, making it well-suited for applications in interactive artistic creation and experiences.

\section{Experiments}
\subsection{Experimental Setup}
\subsubsection{Dataset} We construct a pseudo-aligned tri-modal dataset, ArtiCaps, containing 443,662 image-text-audio samples. Each sample includes i) an image and emotional commentary from ArtEmis dataset , and ii) its most semantically similar musical description with audio sample from MusicCaps dataset, matched via TinyBERT-based semantic similarity of text. Since multiple visual-text samples may align with the same audio clip, the dataset is inherently one-to-many. 
The dataset is split 8:1:1 for training, validation, and testing. To reduce training costs, only 50,000 samples are used for Mel-spectrogram generation stage, with the subset retaining representative semantic and spectral distributions. The HiFi-GAN vocoder is trained on the entire MusicCaps dataset (4,717 samples), using both synthesized and real Mel-spectrograms to enhance generalizability and robustness.

\subsubsection{Hyperparameters} All models are trained on a single NVIDIA RTX 4060 GPU. We adopt the Adam optimizer with a learning rate of $1 \times 10^{-4}$ and a batch size of 32 for 10 epochs. The multimodal fusion module employs a Gated Residual Projector with 1024-dimensional input and 512-dimensional hidden size. The Mel-spectrogram decoder is a 4-layer bidirectional LSTM with 512 hidden units per layer, outputting an 80-dimensional Mel-spectrogram. Training is supervised by a frequency-weighted L1 loss. The model with the lowest validation loss is selected for final evaluation.

\subsubsection{Evaluation Metrics} To comprehensively evaluate the generated audio and the effectiveness of the model, we adopt three categories of evaluation metrics, combining objective computation with perceptual assessment:
\begin{itemize}
    \item \textbf{Perceptual Naturalness: }Fréchet Audio Distance (FAD) is employed to assess the perceptual similarity between the generated audio and referred audio at the distributional level. This metric effectively reflects the naturalness and clarity of the generated samples. 
    \item  \textbf{Structural Consistency:} Mel-Cepstral Distortion (MCD) and Log-Spectral Distance (LSD) quantify the structural alignment between the generated and reference spectrograms. These metrics evaluate the fidelity of temporal and frequency-domain structure.  
    \item  \textbf{Feeling Consistency:} This metric is used to evaluate the consistency between the generated audio and the reference audio. We compute the cosine similarity between their embeddings obtained from the Art2Music model. Although no explicit cross-modal evaluation is conducted, feeling alignment is implicitly ensured during data construction by matching the textual representations of independent image-text and audio-text datasets using TinyBERT. Therefore, the similarity score indirectly reflects the model’s ability to preserve cross-modal feeling conditions. 
\end{itemize}

\begin{figure*}[!t]
\centering
\includegraphics[width=\linewidth]{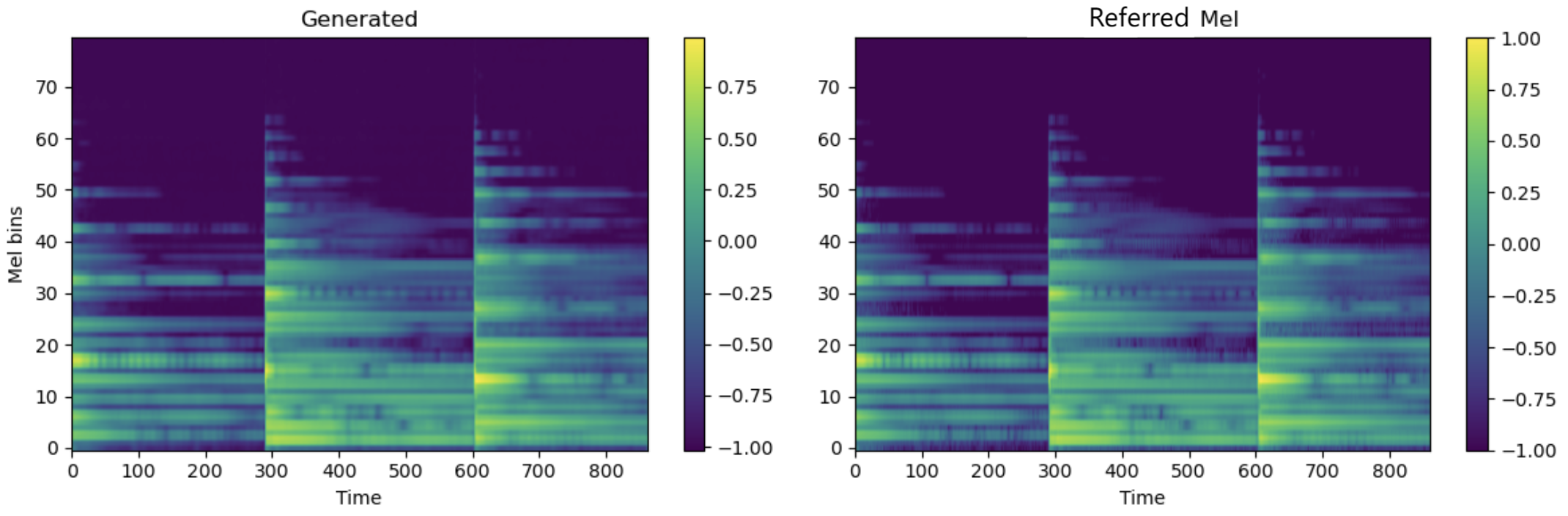}
\caption{Mel-spectrogram comparison between generated audio (left) and referred audio (right). The generated spectrogram preserves overall spectral contour and harmonic structures, demonstrating strong consistency in time-frequency representation.}
\label{fig:mel_comparison}
\end{figure*}

\subsubsection{Input Modality Evaluation}  To evaluate the effectiveness of the model in fusing image and text, we design four input configurations for comparative experiments: 
\begin{itemize}
    \item \textbf{Full Multimodal Input:} The primary setting, where both image and text modalities are provided to the model as complete semantic conditions. 
    \item \textbf{Text-only Input:} Only textual descriptions are used; the image is excluded. 
    \item \textbf{Image-only Input:} Only the visual modality is retained; the textual modality is omitted. 
    \item \textbf{Random Values Input as Baseline:} Gaussian noise vectors are fed as input to simulate unconditioned generation.
\end{itemize}
Moreover, to the best of our knowledge, there are currently no publicly available lightweight multimodal models tailored for the artistic domain. Therefore, direct model-level baselines are not included, and we instead focus on ablation comparisons and the random input setting as reference baselines. For each configuration, we perform a systematic analysis of the generated audio samples across three evaluation dimensions: perceptual naturalness, structural consistency, and semantic consistency. 

\subsection{Results and Discussion}
\begin{table}[!t]
\footnotesize 
\setlength{\tabcolsep}{3pt}
\centering
\caption{Performance under different input configurations. Lower is better ($\downarrow$) for MCD, FAD, and LSD; higher is better ($\uparrow$) for Cosine Similarity.}
\label{tab:ablation-results}
\begin{tabular}{lcccc}
\toprule
\textbf{Model} & \textbf{MCD $\downarrow$} & \textbf{FAD $\downarrow$} & \textbf{LSD $\downarrow$} & \textbf{Cosine Similarity $\uparrow$} \\
\midrule
Baseline         & 13.94 & 0.83 & 11.61 & 0.37 \\
w/o Text         & 12.64 & 0.81 & 10.57 & 0.32 \\
w/o Image        & 12.87 & 0.75 & 18.51 & 0.42 \\
Full Multimodal Input  & \textbf{11.36} & \textbf{0.70} & \textbf{9.64} & \textbf{0.56} \\
\bottomrule
\end{tabular}
\end{table}

Table~\ref{tab:ablation-results} presents the quantitative evaluation results under four input configurations. The full multimodal input configuration achieves the best performance across all metrics, demonstrating the effectiveness in multimodal feeling modeling. Specifically, it achieves the lowest results on \textbf{MCD} (11.36), \textbf{FAD} (0.70), and \textbf{LSD} (9.64), indicating high perceptual realism and spectral fidelity. Meanwhile, the highest cosine similarity score (0.56) achieved under the full multimodal setting shows a stronger feeling alignment with reference audio. Removing either modality results in noticeable performance degradation. In the \textit{w/o Text} setting, \textbf{MCD} increases by approximately 1.28, and the cosine similarity score drops significantly from 0.56 to 0.32. This suggests that text is crucial for semantic anchoring, guiding the overall content of the generated audio. In comparison, removing the image modality (\textit{w/o Image})  preserves some feeling alignment (cosine similarity 0.42) but suffers a dramatic increase in \textbf{LSD} (18.51), highlighting the image modality’s role in maintaining fine-grained spectral structure and time-frequency coherence. As expected, the random input \textit{Baseline} setting performs the worst on all evaluation metrics. This result highlights that the absence of semantic conditioning severely impairs the quality and consistency of generated audio, further confirming the importance of structured semantic input in multimodal generation tasks. It also indirectly validates the effectiveness of our weakly aligned feeling alignment strategy during training.

In summary, our ablation analysis reveals that the image and text modalities contribute complementary information in the Art2Music framework: textual input enhances the accuracy of semantic expression, while visual input contributes to the richness of spectral structure. Their joint use is critical for producing natural, coherent, and semantically consistent audio.

In addition to the quantitative evaluation, we further conduct visual comparisons between the generated and ground-truth spectrograms to assess the model’s capability in structural reconstruction and detail preservation. As shown in Fig.~\ref{fig:mel_comparison}, the Mel-spectrogram generated by Art2Music exhibits high consistency with the referred audio in terms of overall contour, frequency band distribution, and energy patterns. This model effectively retains harmonic features and temporal variations. These results indicate that the proposed model can synthesize structurally aligned and perceptually natural audio under cross-modal conditions, further supporting the conclusions drawn from the quantitative experiments.

\subsection{LLM-based multimodal feeling consistency rating (Case Study)}

\begin{table}[!t]
\scriptsize
\centering
\setlength{\tabcolsep}{3pt}
\caption{The first sample(S1) with left-column image, middle-column textual inputs, and right-column generated audio (icon with link).}
\label{tab:sample_one}
\begin{tabular}{@{}p{2.5cm} p{3.8cm} p{2.0cm}@{}}
\toprule
\multirow{5}{*}{\raisebox{-1.0\height}{\includegraphics[width=2.5cm]{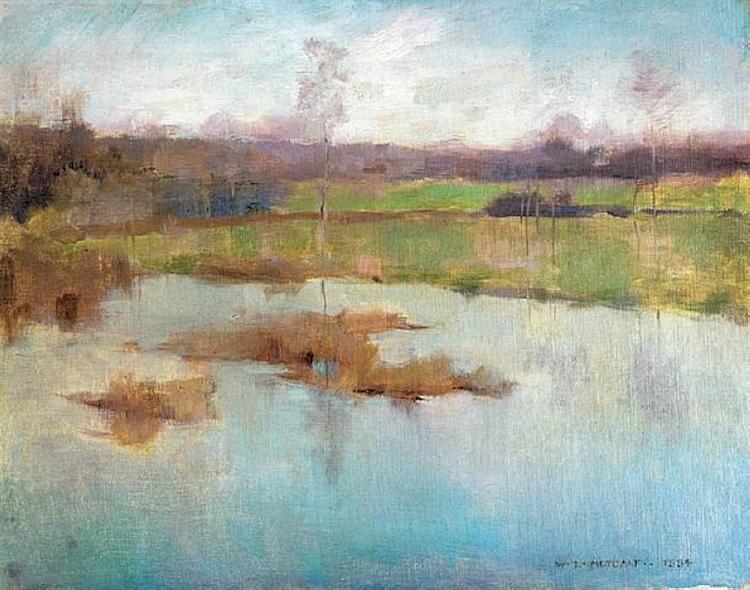}}}
& \textbf{Art Commentary:} The peacefulness of the scene is quite beautiful and I appreciate the watercolor nature.
& \multirow{5}{*}{\raisebox{-1.2\height}{%
    \href{https://github.com/hh-jy/Art2Music/tree/main/LLM-rating_test_samples}
    {\begin{minipage}{2.0cm}
      \centering
      \textbf{Generated Audio}\\[2pt]
      \includegraphics[width=0.7\linewidth]{img/samples/music.png}
    \end{minipage}}
}} \\ 
& \textbf{Painting Emotion:} contentment, appreciate, beautiful \\ 
& \textbf{Audio:} Rapping, Hip hop music \\ 
& \textbf{Audio keywords:} french horns, orchestral piece, warm, relaxing, emotional, slow tempo \\ 
\bottomrule
\end{tabular}
\end{table}

\begin{table}[!t]
\scriptsize
\centering
\setlength{\tabcolsep}{3pt}
\caption{The second sample (S2) with left-column image, middle-column textual inputs, and right-column generated audio (icon with link).}
\label{tab:sample_two}
\begin{tabular}{@{}p{2.5cm} p{3.8cm} p{2.0cm}@{}}
\toprule
\multirow{5}{*}{\raisebox{-.5\height}{\includegraphics[width=2.2cm]{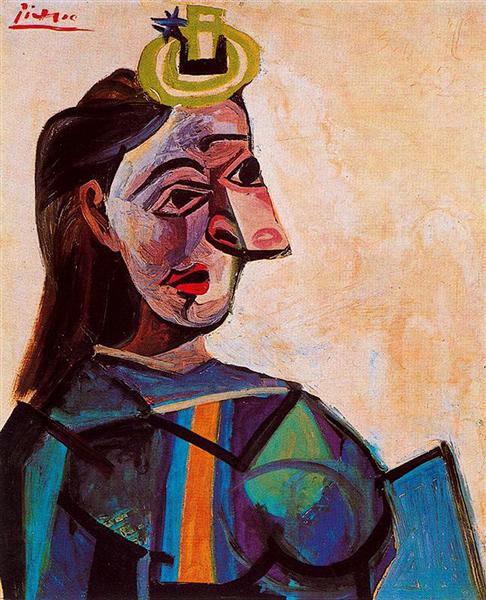}}}
& \textbf{Art Commentary:} The odd proportions on the persons face is quite fun to look at 
& \multirow{5}{*}{\raisebox{-1.3\height}{%
    \href{https://github.com/hh-jy/Art2Music/tree/main/LLM-rating_test_samples}
    {\begin{minipage}{2.0cm}
      \centering
      \textbf{Generated Audio}\\[2pt]
      \includegraphics[width=0.7\linewidth]{img/samples/music.png}
    \end{minipage}}
}} \\ 
& \textbf{Painting Emotion:} amusement,odd, fun \\ 
& \textbf{Audio:} Music,Traditional music \\
& \textbf{Audio keywords:} didgeridoo, live performance, two didgeridoos, low hum, low bass sound, low frequency instrument, wobbly \\
\bottomrule
\end{tabular}
\end{table}

\begin{table}[!t]
\scriptsize
\centering
\setlength{\tabcolsep}{3pt}
\caption{The third sample (S3) with left-column image, middle-column textual inputs, and right-column generated audio (icon with link).}
\label{tab:sample_three}
\begin{tabular}{@{}p{2.5cm} p{3.8cm} p{2.0cm}@{}}
\toprule
\multirow{5}{*}{\raisebox{-.5\height}{\includegraphics[width=1.8cm]{img/samples/maurice-quentin-de-la-tour_study-for-portrait-of-mademoiselle-dangeville.jpg}}}
& \textbf{Art Commentary:} Looks like a painting of some smug weirdo
& \multirow{5}{*}{\raisebox{-1.2\height}{%
    \href{https://github.com/hh-jy/Art2Music/tree/main/LLM-rating_test_samples}
    {\begin{minipage}{2.0cm}
      \centering
      \textbf{Generated Audio}\\[2pt]
      \includegraphics[width=0.7\linewidth]{img/samples/music.png}
    \end{minipage}}
}} \\
& \textbf{Painting Emotion:} smug \\
& \textbf{Audio:} Music,Electronic music,House music,Trance music \\
& \textbf{Audio keywords:} classical music, contemporary, theremin, electronic sounding, instrumental, weird, eerie, eccentric \\
\bottomrule
\end{tabular}
\end{table}

\begin{table}[!t]
\scriptsize
\centering
\setlength{\tabcolsep}{2pt}
\renewcommand{\arraystretch}{1.05}
\caption{Direct multimodal consistency prompt (JSON-only).}
\label{tab:prompt-direct}
\begin{tabularx}{\columnwidth}{@{}>{\ttfamily\raggedright\arraybackslash}X@{}}
\toprule
You are a multimodal feeling consistency evaluator. \\
Given that the audio is generated based on text and images, please determine the similarity between the audio and the text and images in terms of feelings. \\[3pt]
Please output as follows: \textbackslash n \\
1. Give an integer score ranging from 0 to 10. The higher the score, the more consistent the feelings are. \textbackslash n \\
2. Analyze the main feeling keywords of each of the three modalities: audio, text, and images. \textbackslash n \\
3. Explain why you make such a judgment, including but not limited to: \\
- Text: Word choice, sentence structure, artistic conception, etc. \\
- Image: Tone, composition, main elements, etc. \\
- Audio: timbre, rhythm, melody, sense of energy, etc. \\[3pt]
Please ensure that the output strictly complies with the specified JSON Schema. \\
\bottomrule
\end{tabularx}
\end{table}

\begin{table}[!t]
\scriptsize
\centering
\setlength{\tabcolsep}{3pt}
\renewcommand{\arraystretch}{1.1}
\caption{Structured output specification.}
\label{tab:canon-spec}
\begin{tabularx}{\columnwidth}{@{}l l l >{\raggedright\arraybackslash}X@{}}
\toprule
\textbf{Field (dot path)} & \textbf{Type} & \textbf{Constraints} & \textbf{Description} \\
\midrule
scores.Feeling\_alignment & int & $[0,10]$ & Feeling alignment score (higher = better). \\
\midrule
keywords.text  & list\textless str\textgreater & length $1$–$5$ & Main feeling keywords from \emph{text}. \\
keywords.image & list\textless str\textgreater & length $1$–$5$ & Main feeling keywords from \emph{image}. \\
keywords.audio & list\textless str\textgreater & length $1$–$5$ & Main feeling keywords from \emph{audio}. \\
\midrule
explanations.text   & string & required & Text feeling analysis (wording, syntax, imagery). \\
explanations.image  & string & required & Image feeling analysis (tone, composition, elements). \\
explanations.audio  & string & required & Audio feeling analysis (timbre, rhythm, melody, energy). \\
explanations.overall& string & required & Integrated conclusion and rationale across modalities. \\
\midrule
notes & string & optional & Additional remarks (e.g., caveats or edge cases). \\
\bottomrule
\end{tabularx}
\end{table}

Given the subjectivity of feeling perception in music and visual art, and to complement the limitations of objective metrics such as \textbf{FAD}, \textbf{MCD}, \textbf{LSD} and semantic similarity of text, we conduct a small-scale case study ($n=3$; shown in Table~\ref{tab:sample_one},\ref{tab:sample_two},\ref{tab:sample_three}) using an LLM with Gemini~\cite{b23} in direct multimodal mode (image+text+audio). In this setting, the model provides an overall feeling consistency score (0–10) along with modality-specific explanations for text, image, and audio, thereby not only quantifying multimodal feeling alignment but also revealing the underlying reasons for matches or mismatches across modalities in an interpretable manner. The exact prompt of LLM is included in Table~\ref{tab:prompt-direct}, and the outputs follow a structured schema format, with full details provided in the Table~\ref{tab:canon-spec}.

As shown in Appendix~\ref{sec:LLM_outputs}, which presents the LLM outputs, all three examples achieve relatively high overall scores, indicating that the proposed framework attains good feeling alignment across modalities. The score differences mainly stem from subtle variations in expressive detail: for instance, \textbf{S1} exhibits strong consistency around the theme of peacefulness, whereas \textbf{S2} and \textbf{S3} receive slightly lower scores due to minor feeling mismatches (e.g., playful amusement vs. the primal timbre of the didgeridoo, or smug self-satisfaction vs. an eerie atmosphere). These findings highlight the subjectivity of cross-modal feeling perception while reinforcing the robustness of our overall conclusion that the framework effectively captures and aligns feelings at a macro level.

\section{Conclusion}
This study proposes Art2Music, a lightweight and feeling-aligned multimodal audio generation framework. It synthesizes perceptually natural and feeling-aligned audio from artistic images and their textual descriptions. Addressing the lack of aligned tri-modal data, we construct a pseudo-aligned dataset using text-based semantic matching and propose a two-stage framework: image and text features are encoded by OpenCLIP and fused through a gated residual module, decoded into Mel-spectrograms with a lightweight sequence model, and reconstructed into audio with a HiFi-GAN vocoder. Experimental results demonstrate the effectiveness of our approach across multiple evaluation dimensions. The proposed framework consistently outperforms ablation settings and a random baseline in perceptual quality (FAD), structural fidelity (MCD, LSD), and feeling consistency. Our analysis reveals that textual input enhances semantic expressiveness, while visual content contributes to spectral structure, highlighting the complementary roles of both modalities. Additionally, the model maintains robust performance under data-limited conditions, underscoring its practicality in low-resource creative scenarios. Furthermore, an LLM-based case study corroborates these findings, showing consistently high feeling alignment scores across modalities, with only minor variations attributable to subjective feeling perception.

Despite its promising results, Art2Music faces two limitations: 1) it relies on weakly aligned text-based semantic matching to construct pseudo-aligned data. However, it still lacks systematic verification of cross-modal subjective consistency; and 2) the diversity and contextual adaptability of the generated audio are still constrained by the limited music dataset (i.e., MusicCaps) resources. Art2Music opens new possibilities for artistic and creative applications, such as personalized exhibition soundtracks, interactive installation soundscapes, and automated music generation for visual artworks. Future work will focus on enhancing cross-modal alignment through advanced retrieval and supervision strategies, enriching stylistic diversity (e.g., multi-layered soundscapes and emotionally rich ambiances), and incorporating user feedback to improve contextual adaptability and controllability. 

\appendix

\subsection{LLM Outputs}
\label{sec:LLM_outputs}
Table~\ref{tab:gemini_case_output} shows the output of the LLM.

\begin{table*}[!t]
\centering
\caption{Gemini-based multimodal feeling consistency ratings (0--10 scale) with modality-specific explanations.}
\label{tab:gemini_case_output}
\footnotesize
\setlength{\tabcolsep}{4pt}
\renewcommand{\arraystretch}{1.15}
\begin{tabularx}{\textwidth}{c c >{\raggedright\arraybackslash}p{0.26\textwidth} >{\raggedright\arraybackslash}X >{\raggedright\arraybackslash}X}
\toprule
\textbf{Sample}&  \textbf{Score}  &\textbf{Feeling Keywords} (from Text, Image, Audio)&\textbf{Explaination (Text, Image, Audio)} &\textbf{Explaination (Overall)}\\
\midrule
\textbf{S1}& 9 &\textbf{Text}: peacefulness, beautiful, appreciate, watercolor \newline
\textbf{Image}: serene, tranquil, landscape, watercolor, calm \newline
\textbf{Audio}: warm, relaxing, orchestral, slow tempo, emotional& \textbf{Text}: "The text directly states peacefulness and describes the scene as quite beautiful, expressing an appreciation for its watercolor nature. These words evoke a gentle, calm, and aesthetically pleasing sentiment, suggesting a quiet contemplation and enjoyment of the scene." \newline
\textbf{Image}: "The image, a watercolor painting of a landscape, features soft, muted tones of blue, green, and brown, creating a tranquil and ethereal atmosphere. The broad, gentle brushstrokes characteristic of watercolor, combined with the reflective water and open field composition, contribute to a pervasive sense of serenity and calm beauty." \newline
\textbf{Audio}: "The audio is an orchestral piece characterized by the warm, resonant tones of french horns and a slow tempo. This combination generates a relaxing, contemplative, and slightly emotional soundscape. The absence of sharp rhythms or dissonant harmonies fosters a feeling of peacefulness and gentle introspection."&"All three modalities converge on a strong theme of peacefulness, beauty, and gentle appreciation. The text explicitly states these feelings, the image visually embodies them through its soft aesthetic and tranquil composition, and the audio reinforces them with its warm, slow, and orchestral character. The consistency in conveying a serene and contemplative mood is very high."\\
\midrule
\textbf{S2}& 8 &\textbf{Text}: amusement, odd, fun \newline
\textbf{Image}: distorted, abstract, unconventional, colorful, intriguing \newline
\textbf{Audio}: didgeridoo, wobbly, unusual, droning, primal& \textbf{Text}: "The text directly uses words like odd and fun to describe the proportions of the face, conveying a feeling of amusement and lighthearted appreciation for the unconventional aspects. The phrase fun to look at explicitly sets a playful and positive tone." \newline
\textbf{Image}: "The Picasso painting, characterized by its Cubist style, features highly distorted and fragmented facial features and body. The bold, contrasting colors (blues, greens, oranges, purples) and the abstract composition create an unconventional and intriguing visual experience. This distortion directly aligns with the odd aspect mentioned in the text, inviting a sense of curiosity and amusement." \newline
\textbf{Audio}: "The audio features the unique, low-frequency sound of a didgeridoo. The wobbly and low hum characteristics of the sound, as described by the keywords, contribute to an unusual and somewhat primal or earthy feeling. While not overtly amusing, the distinct and unconventional nature of the didgeridoo's timbre and rhythm provides an auditory parallel to the odd and unconventional qualities found in the text and image."&"The three modalities demonstrate a strong consistency in exploring the theme of the unconventional or odd in a positive, intriguing, and somewhat playful manner. The text explicitly states that the odd proportions are fun, which the distorted Cubist image perfectly embodies through its abstract and unconventional representation. The audio, with its unique, wobbly, and droning didgeridoo sounds, offers an auditory representation of something unusual and captivating, mirroring the spirit of amused appreciation for the unconventional present in both the textual description and the visual art."\\
\midrule
\textbf{S3}& 6 &\textbf{Text}: smug, weirdo \newline
\textbf{Image}: smug, self-satisfied, enigmatic, unsettling \newline
\textbf{Audio}: eerie, weird, eccentric, unsettling, mysterious& \textbf{Text}: "The text uses strong descriptive words like 'smug' and 'weirdo' to immediately establish a feeling of disdain or critical observation towards the subject, implying an unconventional and self-satisfied character." \newline
\textbf{Image}: "The image, a charcoal drawing, depicts a person with a subtle, closed-lip smile and a gaze that suggests self-satisfaction or amusement. This expression clearly conveys 'smugness'. The overall composition and style give the subject an enigmatic quality that can be interpreted as 'weird' or unusual." \newline
\textbf{Audio}: "The audio features a distinctive, high-pitched, and wavering electronic sound, strongly reminiscent of a theremin. Its slow, sustained nature, lack of traditional melody, and unconventional timbre create a pervasive sense of eeriness, weirdness, and eccentricity, evoking an unsettling and mysterious atmosphere."&"There is a moderate alignment in feelings across the three modalities. Both the text and image prominently feature 'smugness' and a sense of 'weirdness' or 'unsettling' peculiarity. The audio successfully captures the 'weird', 'eerie', and 'unsettling' aspects with its unusual electronic sounds, creating an atmosphere of mystery and strangeness. However, the confident, self-satisfied 'smugness' that is central to the image's expression and the text's description is not directly conveyed by the audio, which focuses more on atmospheric unease rather than a specific emotion of superiority or self-satisfaction. The common thread is the feeling of something unusual and slightly unsettling."\\
\bottomrule
\end{tabularx}
\end{table*}

\clearpage
\section*{Authors’ background}
\begin{table}[!htbp]
\centering
\renewcommand{\arraystretch}{1.3}
\begin{tabular}{|p{3cm}|p{3cm}|p{3.5cm}|p{3.5cm}|p{3.5cm}|}
\hline
\textbf{Name} & \textbf{Prefix} & \textbf{Research Field} & \textbf{Email} & \textbf{Personal website} \\
\hline
 Jiaying Hong& Master Student& Multimodal Models, NLP, Deep Learning, LLM& hongjalynn@gmail.com& -\\
\hline
 Ting Zhu & PhD student & Expressive speech synthesis, Conversational AI & t.zhu11@newcastle.ac.uk & -\\
\hline
Thanet Markchom & Postdoctoral Research Assistant & Computer Vision, NLP & thanet.markchom.reading.ac.uk & -  \\
\hline
 Huizhi Liang& Senior Lecturer& Data Mining, Machine Learning, Natural Language Processing, Recommender Systems, Personalisation& huizhi.liang@newcastle.ac.uk& https://ellyliang.com/\\
\hline
\end{tabular}
\end{table}

\end{document}